# La motivation au pied de la lettre

Construction et expression des aspirations scolaires sur Parcoursup

## Résumé

Cet article analyse les modalités de l'encadrement et de l'expression des aspirations des lycéens français. Il apporte un éclairage nouveau sur les inégalités de parcours entre filière générale versus filière technologique et professionnelle. À travers l'analyse d'une enquête nationale et d'un corpus de lettres de motivation rédigées par des candidats à une licence de sociologie, il montre que, faute de moyens, les enseignants disposent principalement de deux types de stratégies d'accompagnement à l'orientation.

Les enseignants ont pour habitude de cibler et de concentrer leur soutien sur les « bons élèves » dans les filières technologiques et professionnelles, tandis que, dans les filières générales, ils délèguent certaines étapes des procédures de suivi aux familles. Ces différentes stratégies ont des effets sur la manière dont les lycéens intériorisent les prescriptions scolaires et les restituent dans des lettres de motivation. Par le soutien resserré dont ils bénéficient auprès des enseignants, les « bons élèves » des filières technologiques et professionnelles intériorisent fortement les consignes et leur place dans la hiérarchie scolaire. Dans les filières générales, l'expression des aspirations des étudiants est beaucoup plus dépendante de leur capital familial.

## Autrices

**Marie-Paule Couto**
Maîtresse de conférences en sociologie, Université Paris 8 Vincennes-Saint-Denis et Cresppa-CSU.
Thèmes de recherche : Accès à l'enseignement supérieur, trajectoires étudiantes et socialisation universitaire.
mp.couto@outlook.fr

**Marion Valarcher**
Doctorante en sociologie et ATER, Sciences Po, OSC et Université de Paris
Thèmes de recherche : Accès à l'enseignement supérieur, Projets d'orientation post-bac, trajectoires, socialisation et représentations des lycéens et lycéennes.
marion.valarcher@sciencespo.fr

# Introduction

Les « guichets » de l'enseignement supérieur ont connu ces dernières décennies une série de transformations communes à d'autres organismes publics, celles-ci visent à gérer et rationaliser les flux de candidats à l'entrée du supérieur (Lemêtre & Orange, 2017) comme ceux des patients vers l'hôpital (Belorgey, 2010) ou d'administrés aux guichets de la Caisse d'allocations familiales (Deville, 2018)[1]. Cette tendance s'est traduite notamment par une dématérialisation des procédures affectant l'accès aux droits des usagers ou la répartition des élèves dans « l'espace hiérarchisé de l'enseignement supérieur » (Convert, 2003). Du système informatisé Ravel[2] (1990-2008) à APB[3] (2009-2017), les études pointent le rôle des réformes institutionnelles dans l'inégale répartition des élèves dans les formations de l'enseignement supérieur en particulier selon leur niveau scolaire et leur origine sociale (Frouillou 2017 ; Frouillou et al., 2020). Si l'inégale distribution des élèves dans l'espace de l'enseignement supérieur est établie de longue date (Duru Bellat & Kieffer, 2008), la mise en place en 2018 de la nouvelle plateforme d'affectation Parcoursup – et avec elle l'extension de la sélection sur dossier scolaire à l'ensemble des formations de l'enseignement supérieur– a ravivé les débats sur la ségrégation sociale et scolaire au sein du système éducatif français.

En 2010, Verley et Zilloniz (2010) distinguaient déjà trois secteurs d'enseignements : un segment sélectif recrutant les élèves les mieux armés scolairement et socialement composé notamment des classes préparatoires aux grandes écoles (CPGE), un segment professionnalisant, « sélectionnant un petit nombre d'étudiants, qui ne sont pas des héritiers, en vue de les préparer à une insertion professionnelle à court terme (STS[4], IUT[5]) » et enfin un segment universitaire « ouvert » accueillant un public diversifié (Verley & Zilloniz, 2010, p. 6) ; le recrutement social de chaque segment tenant à la fois au tri scolaire opéré à leur entrée et aux aspirations inégales des lycéens. À la suite de l'élargissement du tri scolaire au secteur universitaire, les travaux se sont principalement centrés sur les opérations de classement des candidatures à l'entrée des formations (Chauvel et al., 2020). Pourtant, d'après une évaluation récente de l'Insee, « une grande partie de la ségrégation à l'entrée de l'enseignement supérieur est déjà présente dans les vœux exprimés par les candidats, alors que les classements des candidats opérés par les formations [à l'exception notable de certaines licences en tension] n'y contribuent que faiblement » (Bechichi et al., 2021, p. 105).

---

[1] Nous remercions particulièrement Alice Olivier, Mathieu Ferry et les relecteurs anonymes de la revue OSP pour leurs relectures attentives. Merci également à Joanie Cayouette-Remblière et à Lucile Bouré pour leurs suggestions.

[2] Système de sectorisation pour 14 licences en Île-de-France

[3] Admission post bac

[4] Section de technicien supérieur

[5] Institut Universitaire de Technologie

La présente recherche s'intéresse à la production des vœux ou des aspirations scolaires en amont du travail réalisé par les commissions de sélection dans un contexte renouvelé par la réforme de l'accès à l'enseignement supérieur. La loi Orientation et Réussite des Étudiants (ORE) a également modifié leurs modalités d'expression sur le portail Parcoursup : désormais, la procédure impose à tous les candidats la composition d'un dossier comprenant un CV et un projet de formation motivé. Propre à chaque vœu, le projet motivé est officiellement présenté comme « une expression personnelle » visant à informer les responsables de formations de l'enseignement supérieur « sur la motivation ou les souhaits du lycéen »[6]. Pour accompagner les lycéens dans leur choix et dans la composition de ce dossier, la réforme prévoit un accompagnement renforcé des élèves par deux professeurs principaux et des moments dédiés à l'orientation vers l'enseignement supérieur au lycée. Dans la lignée des travaux sur la « canalisation » des aspirations scolaires (van Zanten, 2015, p. 81), cet article entend interroger les modalités d'encadrement et d'expression des aspirations dans ce cadre en procédant à l'analyse croisée de l'Enquête nationale sur la « Transition du Secondaire au Supérieur » (TSS) de l'Observatoire de la Vie Étudiante (OVE) et d'un corpus de projets motivés rédigés par des lycéens (voir *infra*).

Différents acteurs contribuent à définir les aspirations, à commencer par « l'École [qui] participe activement à la production des « vœux » qui ne sont qu'en apparence formulés de manière libre et autonome par l'élève et sa famille » (Cayouette-Remblière, 2014, p. 62). Le rôle de l'encadrement et des jugements scolaires dans la construction des aspirations, en particulier auprès des élèves les moins armés socialement et scolairement a été mis en évidence avant la réforme. L'École oriente les lycéens vers des cursus qui leurs seraient dédiés – à travers les avis du conseil de classe, les conseils des enseignants, la documentation mobilisée, etc. – notamment les bacheliers professionnels vers les sections de techniciens supérieurs (STS) (Orange, 2009). En prédéfinissant des « filiations naturelles entre formations secondaires et supérieures », elle pèse sur les projections scolaires (Orange, 2010, p. 35). Convert (2010) décrit ainsi « deux types de trajectoires scolaires et […] expériences de la transition secondaire- supérieur radicalement différents », l'un relatif aux bacheliers généraux et l'autre aux lycéens technologiques encouragés à s'inscrire dans le segment professionnalisant de l'enseignement supérieur (Convert, 2010, p. 16). Les données nationales sur la procédure d'affectation 2018 montrent en effet que les candidats en terminale générale formulent davantage de souhaits et que leurs préférences se portent majoritairement sur les licences. «

[6] https://cache.media.eduscol.education.fr/file/Parcoursup_2020/51/9/Guide-pratique-Parcoursup-2020accompagner-les-lyceens_1223519.pdf

Comparativement, [les élèves] en terminale technologique et professionnelle demandent un nombre plus réduit de formations [et] préférentiellement en STS, celles-ci représentant respectivement 54 % et 84 % de leurs vœux ». (DEPP, 2019, p. 190). Nous formulons l'hypothèse que l'encadrement par l'École ne prend pas la même forme ou ne s'exerce pas avec la même force selon les types de classes fréquentées (générales, technologiques ou professionnelles).

L'origine sociale influence également la manière dont les élèves et leur famille appréhendent les prescriptions de l'institution scolaire. Les enfants de classes populaires « moins familiarisés et préparés à l'enseignement supérieur par leurs parents sont plus enclins que d'autres, car plus dépendants, à recevoir et accepter » les injonctions de l'école concernant leur cursus (Orange, 2010, p. 39). L'École intervient également dans la modération des ambitions des lycéens ruraux ou encore des femmes (Blanchard et al., 2016 ; Lemêtre & Orange, 2016). Truong (2013) détaille à l'inverse comment il s'est attaché – dans le cadre d'une ethnographie participante – à réassurer certains élèves de milieux populaires : disposant de ressources limitées, l'enseignant s'adapte en concentrant ses « efforts sur un petit nombre d'élèves dont [il pense] qu'ils sont influençables et dont le dossier puisse retenir l'attention d'une filière de type sélectif » (Truong, 2013). Nous supposons que l'encadrement varie également selon les profils socio-scolaires des élèves (saisie par le statut de boursier et la mention au bac) et considérons que ces différences contribuent à l'intériorisation par les lycéens de leur place dans les hiérarchies scolaires.

La littérature a également mis en évidence que le recrutement social des enseignants avait des effets sur leurs pratiques. Les manières de s'investir auprès des élèves dépendent notamment des modalités d'entrée dans le métier (Llobet, 2011). Or, sur certains aspects les enseignants des voies professionnelles se distinguent de ceux exerçant en lycée général. Plus souvent professionnels reconvertis dans l'enseignement que leurs homologues des filières générales, ils justifient leurs reconversions par une quête de sens et une volonté d'accompagner des jeunes (Dozolme & Ria, 2019 ; Jellab, 2009 ; Tanguy, 1991). D'origines plus modestes que leurs collègues des lycées GT (Général et Technologique), ils promeuvent « les valeurs de travail et de mérite qui [leur] ont permis de grimper dans l'échelle sociale » (CayouetteRemblière, 2013, p. 154). En outre, la procédure d'orientation est davantage encadrée dans les voies professionnelles qu'en terminale générale (Lemêtre & Orange, 2017).

En comparant les pratiques au sein d'établissements au recrutement social distinct, van Zanten (2015) souligne en effet des degrés d'investissement variés en lien avec les priorités des équipes pédagogiques. Dans les lycées favorisés, le travail d'information à l'orientation est « délégué […] aux élèves et leurs familles qu'on présume capables de mobiliser un ensemble de ressources extérieures […] pour éclairer leurs jugements » (van Zanten, 2015, p. 89). La mise en œuvre de la réforme de l'accès à l'enseignement supérieur est cependant contrainte par les obstacles matériels auxquels font face les enseignants dans les lycées considérés (Daverne-Bailly & Bobineau, 2020). En outre, selon Bodin et Orange (2019, p. 218), le nouveau dispositif contribue à « dépister des lycéens déviants », dont les projets sont jugés incohérents. Il s'agit de pousser les bacheliers à formuler des vœux « adéquats » (Frouillou et al., 2020). Pourtant, selon le gouvernement, le nouveau portail est censé permettre aux élèves d'exprimer librement leurs préférences. Plusieurs changements introduits par Parcoursup accentuent en effet l'idéologie du « libre choix » des élèves déjà présente dans le portail APB (Couto et al., 2021). En mettant fin à la hiérarchisation des vœux – cruciale sur l'ancien portail – le fonctionnement de Parcoursup doit notamment limiter les mécanismes d'autocensure[7]. Nous faisons l'hypothèse que les variations dans l'encadrement des élèves sont susceptibles d'influencer l'expression de leurs projections scolaires sur Parcoursup.

L'application de la loi ORE est donc traversée par plusieurs tensions. D'abord, « tout en prônant le libre choix des élèves, les pouvoirs publics ont multiplié les dispositifs visant à encadrer les aspirations des bacheliers » (Couto et al., 2021, p. 25). Ensuite, elle prévoit un accompagnement individualisé des élèves que les enseignants peinent à appliquer par manque de ressources. Il s'agit ici de comprendre comment les étudiants ont construit leur orientation dans ce contexte et comment cela se traduit dans l'expression de leurs aspirations sur Parcoursup.

Cet article se propose de répondre à trois objectifs. Le premier est d'étudier l'effet socialisateur de la série du baccalauréat sur la structuration et l'expression des aspirations scolaires. Plus précisément, il s'agit de décrire de quelle façon la filière de baccalauréat constitue un espace d'intériorisation des hiérarchies scolaires. Le second est de mettre en évidence le rôle que peut jouer l'école sur l'expression de ces aspirations selon le bac d'origine et les résultats scolaires. Enfin, il s'agit de documenter la place des ressources familiales dans ces mêmes expressions selon les différentes filières.

# Méthodologie

Cet article s'intéresse à la manière dont les lycéens construisent et expriment leurs aspirations scolaires à partir de deux matériaux originaux, d'une part l'Enquête nationale sur la Transition du Secondaire au Supérieur (TSS) de l'Observatoire de la Vie Étudiante (OVE) et d'autre part un corpus de lettres de motivation rédigées par des lycéens dans le cadre de leur candidature à une licence de sociologie en Île-de-France.

## Matériel

En questionnant les étudiants inscrits à l'université, en IUT et en CPGE sur la manière dont s'est déroulée leur orientation durant leur année de terminale, l'enquête TSS permet de décrire le poids de l'École dans la structuration des aspirations scolaires. Elle interroge les élèves sur leurs vœux favoris en 2018, les motivations à l'origine de ces vœux ainsi que sur l'accompagnement assuré par leurs enseignants ou leurs proches à différentes étapes de la procédure (Belghith et al., 2019). Certaines questions portent spécifiquement sur la rédaction des CV et des projets de formation motivés. Officiellement, le projet motivé (d'au maximum 1500 caractères) équivaut à une lettre de motivation. En 2018, les consignes fournies aux lycéens pour rédiger ces lettres les encourageaient à y faire figurer leur « projet professionnel » ou, à défaut, le domaine d'activité qui intéresse ; elles conseillaient également de présenter certaines des qualités acquises dans le cadre d'activités extrascolaires ou de stages. Enfin, les documents officiels recommandent d'expliciter « les démarches effectuées pour se renseigner sur la formation ».

[7] D'après la documentation officielle, « le fait de hiérarchiser les vœux [sur la plateforme APB] obligeait les lycéens à préjuger des filières dans lesquelles ils seraient acceptés, ce qui conduisait à l'élaboration de stratégies souvent contreproductives et nourrissait l'autocensure » (MESRI- MEN, 2018, p. 12).

En complément de l'exploitation de l'enquête TSS, nous avons procédé à l'analyse de l'ensemble des lettres de motivation envoyées par les 743 élèves de terminale qui ont candidaté dans une licence de sociologie en 2018[8]. Associés aux informations relatives à la trajectoire scolaire (type de baccalauréat, mention au baccalauréat, bulletins scolaires, avis des enseignants sur l'orientation, etc.) et au profil de leurs auteurs (statut de boursier, âge, sexe), ces lettres constituent un moyen original d'appréhender les rapports contrastés des élèves à une même formation. Les données ont été recueillies directement depuis la plateforme Parcoursup en téléchargeant l'intégralité des dossiers de candidatures lors de la phase principale de l'année 2018, dans le cadre d'un accord avec l'établissement. En outre, l'étude de ce corpus dialogue utilement avec les données nationales de l'OVE dans la mesure où elle permet de saisir les aspirations ou projections de lycéens aux profils diversifiés. En effet, contrairement aux filières du « segment sélectif » qui captent principalement les candidatures des lycéens les mieux armés socialement et scolairement, la licence en question reçoit des dossiers d'élèves de lycées GT comme professionnels, appartenant aux classes supérieures comme aux classes populaires. L'analyse des lettres permet donc de saisir l'hétérogénéité des formes d'expression des aspirations de la part d'élèves avec des profils socio-scolaires diversifiés et aspirant à rejoindre des segments différents de l'enseignement supérieur.

## Une méthode mixte d'analyse de matériaux qualitatifs et quantitatifs

L'analyse des deux matériaux s'est faite simultanément et les résultats présentés ici sont issus d'allers-retours fréquents entre les deux sources. Ainsi, le raisonnement par cas sur les lettres a donné à voir des disparités dans les formulations selon certaines caractéristiques sociales des élèves et ces caractéristiques ont ensuite guidé l'exploitation de l'enquête TSS dans un souci de systématisation et généralisation (Aguilera & Chevalier, 2021, pp. 381-382). Symétriquement, les différences dans les pratiques observées dans TSS ont donné des pistes d'interprétation des lettres de motivation, montrant bien la complémentarité des matériaux et de leur analyse (Aguilera & Chevalier, 2021, p. 385).

[8] En plus de l'intégralité des dossiers Parcoursup, nous disposons d'éléments biographiques collectés auprès de certains des étudiants sous la forme de fiches de renseignement distribuées dans un cours de sociologie ou d'entretiens lorsque les candidats ont finalement choisi de s'inscrire dans un autre diplôme. Peu mobilisés dans le texte, ces éléments ont également nourri l'analyse.

Afin de systématiser l'analyse des lettres de motivation, nous avons réalisé une analyse factorielle de correspondances (AFC) ainsi qu'une classification hiérarchique descendante (CHD) à l'aide du logiciel Iramuteq (Ratinaud, 2020). La méthode de l'AFC sur données textuelles (Benzecri, 1982) rend compte graphiquement à la fois de la proximité entre des individus qui mobilisent le même type de vocabulaire dans les lettres mais également de la proximité entre des mots utilisés dans les mêmes lettres. La CHD (Reinert, 1983) propose une division du corpus en différentes classes de mots significativement employés dans les mêmes lettres. Elle permet de faire émerger les structures caractéristiques des lettres de motivation selon les propriétés sociodémographiques des élèves qui les ont rédigés (Garnier & Guérin-Pace, 2010). Sur le graphique (figure 2) présenté ci-après, le logiciel fait apparaître les classes de la CHD sur le plan de l'AFC, dans différentes couleurs. Toutes les variables sociodémographiques (sexe, type de baccalauréat, mention au baccalauréat, statut de boursier, etc.), sont utilisées comme des variables supplémentaires : elles permettent de décrire les caractéristiques sociales et scolaires des auteurs des différents groupes, mais ne participent pas à donner sa structure au graphique de l'AFC, ni à constituer les classes de la CHD qui résultent uniquement des proximités dans les usages du vocabulaire. Cette méthode inductive permet de déconstruire certaines catégories institutionnelles en proposant des regroupements que le chercheur n'avait pas envisagés. Dans l'analyse présentée ci-après, c'est le cas des classes regroupant les lettres de bacheliers technologiques et professionnels (pourtant scolarisés dans des établissements différents)[9].

Dans l'exploitation de TSS, tous les pourcentages présentés sont pondérés et sont présentés avec des intervalles de confiance à 90 %. Du fait de la taille de l'échantillon, nous nous autorisons à commenter des intervalles se recoupant dans le cas où un test (Chi-deux) est statistiquement significatif et si les observations sont confortées par le reste des traitements, considérant « la convergence [systématique] des indicateurs comme critère de signification sociologique » (Blanchard et al., 2016, p. 115).

[9] Par souci de lisibilité, nous avons choisi d'agréger ces deux groupes dans la présentation des résultats de l'enquête TSS, ce regroupement ayant aussi pour avantage de pallier les effectifs relativement faibles de bacheliers professionnels.

# Résultats

Pour répondre à nos objectifs, nous présenterons tout d'abord les effets de la série du bac sur la structuration et l'expression des aspirations, puis dans un second temps, le rôle de l'École et des enseignants sur ces mêmes aspirations et enfin, l'importance des ressources familiales dans l'orientation. Dans le cadre de ces trois sous-parties, nous ferons à chaque fois dialoguer l'espace du vocabulaire issu de l'AFC et les données de l'enquête TSS.

## L'intériorisation des hiérarchies scolaires ou l'effet de la série du bac sur la structuration et l'expression des aspirations

Si l'ancienne plateforme d'affectation à l'enseignement supérieur (APB) imposait aux candidats de hiérarchiser leurs vœux de poursuite d'études, ce n'est plus le cas avec Parcoursup. L'enquête TSS interroge néanmoins les étudiants sur leurs trois vœux les plus souhaités ainsi que sur les principales raisons qui ont motivé ces vœux. Rappelons que seuls les étudiants finalement inscrits à l'Université (y compris en IUT) et en CPGE ont été interrogés par l'OVE.

*Des choix d'orientation contrastés selon le type de baccalauréat et les motivations associées*

Comme attendu, les bacheliers technologiques et professionnels montrent une préférence pour les filières courtes et professionnalisantes de l'enseignement supérieur (BTS et DUT) en comparaison aux lycéens des voies générales. Si les étudiants issus des voies technologiques et professionnelles demeurent nombreux à évoquer une licence à l'université comme vœu de prédilection (45 %), cette proportion reste nettement inférieure à celle constatée pour les lycéens généraux (65 %). En outre, les CPGE semblent quasiment absentes de leur champ des possibles, 4 % seulement désignant ce type de formation comme leur vœu le plus souhaité contre 14 % des bacheliers généraux.

Les taux de réussite en fonction du baccalauréat d'origine figurent sur le portail Parcoursup et incitent ainsi les bacheliers technologiques ou professionnels à se diriger vers le segment professionnalisant de l'enseignement supérieur (Bodin & Orange, 2019, p. 219). Au sein de ce segment, les élèves se distribuent ensuite selon leur niveau scolaire (voir 2e partie) et leur origine sociale.

L'enquête TSS permet de considérer les raisons avancées par les étudiants pour justifier leurs inclinaisons. Les étudiants sont nombreux à présenter l'intérêt pour le contenu des études comme à l'origine de leurs vœux, mais ce critère est davantage exprimé par les bacheliers généraux : 51 % présentent cette dimension comme leur principale source de motivation contre 39 % des bacheliers technologiques et professionnels. Si les élèves non-boursiers[10] semblent privilégier cet aspect par rapport aux autres lycéens, ils se distinguent encore de 10 points selon le type de baccalauréat (Figure 1.a). Ce phénomène s'explique car les élèves des séries technologiques et professionnelles – qui occupent une position moins favorable dans la hiérarchie des filières du baccalauréat – se voient contraints de prendre en compte l'adéquation de la formation visée à leur baccalauréat. Quand 27 % des bacheliers technologiques et professionnels considèrent cette adéquation comme à l'origine de leurs choix, seuls 11 % des bacheliers généraux présentent cet aspect comme décisif (Figure 1.b)[11]. La documentation utilisée dans les lycées et accessible en ligne contribuant à « inscrire l'ordre social dans les cerveaux et à convaincre nombre d'élèves de leur place dans l'espace hiérarchisé que constitue [l'enseignement supérieur] », les lycéens technologiques et professionnels intègrent plus que les autres cette contrainte au moment de construire leur orientation (Cayouette-Remblière, 2014, p. 62). L'injonction croissante à la professionnalisation conduit aussi les étudiants à penser leurs choix de poursuite d'études autour d'un projet professionnel (Verley & Zilloniz, 2010), c'est le cas de 44 % d'entre eux dans l'enquête TSS. Si les écarts sont modestes entre les bacheliers des différentes voies, cette injonction reste sensiblement plus forte au sein des séries technologiques et professionnelles qui préparent plus directement à un emploi : 6 points séparent les élèves des séries technologiques ou professionnelles des lauréats d'un bac général, ces derniers ayant davantage à cœur de conserver le plus possible de portes ouvertes (Figures 1.c et 1.d).

[10] Nous utilisons comme variable indicatrice de l'origine sociale le statut de boursier par souci de cohérence entre les différentes sources exploitées. En effet, cette information figure à la fois dans l'enquête TSS et dans les dossiers de candidature à l'entrée de la licence de sociologie considérée. Calculées sur la base du revenu brut global figurant sur l'avis de l'imposition de la famille ainsi que sur les charges de la famille, les bourses sont destinées aux étudiants disposant de ressources familiales limitées. Les résultats de l'enquête TSS ont été éprouvés à l'aide d'autres indicateurs d'origine sociale, tels que la « classe sociale » – construite à partir de la profession du père et à défaut de la mère dans la base de données – et le niveau d'études le plus élevé des parents.

[11] À titre d'exemple, 2 points seulement séparent les lycéens technologiques des bacheliers professionnels qui partagent l'expérience d'une position peu valorisée dans les hiérarchies scolaires ou moins avantageuse que celle des lycéens généraux.

*Figure 1 - Raisons avancées pour justifier les trois premiers vœux favoris sur Parcoursup (en %)*

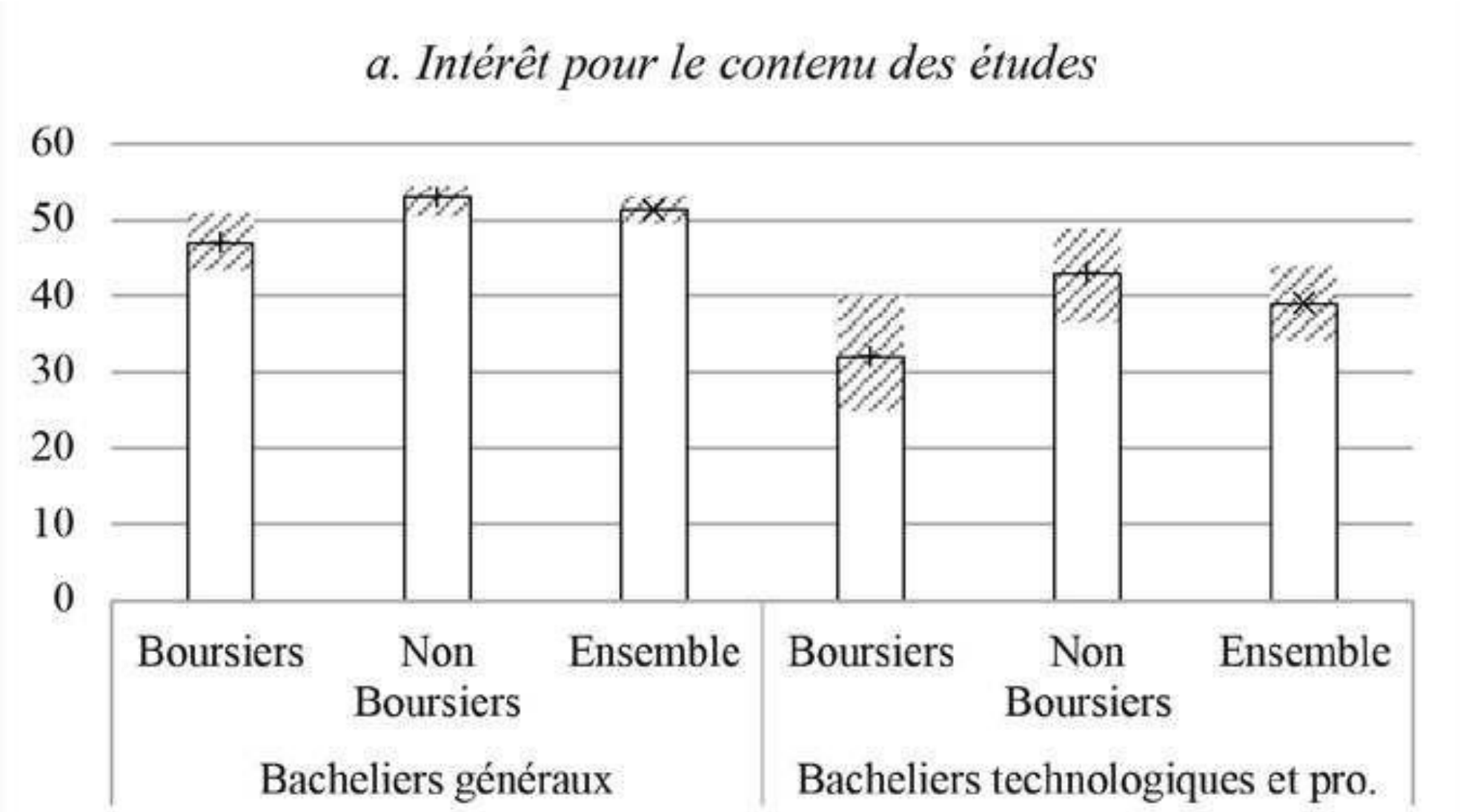

+ Khi2 de Rao-Scott : 23 ddl : 3 p < 0,01
× Khi2 de Rao-Scott : 48,8 ddl : 1 p <0,01

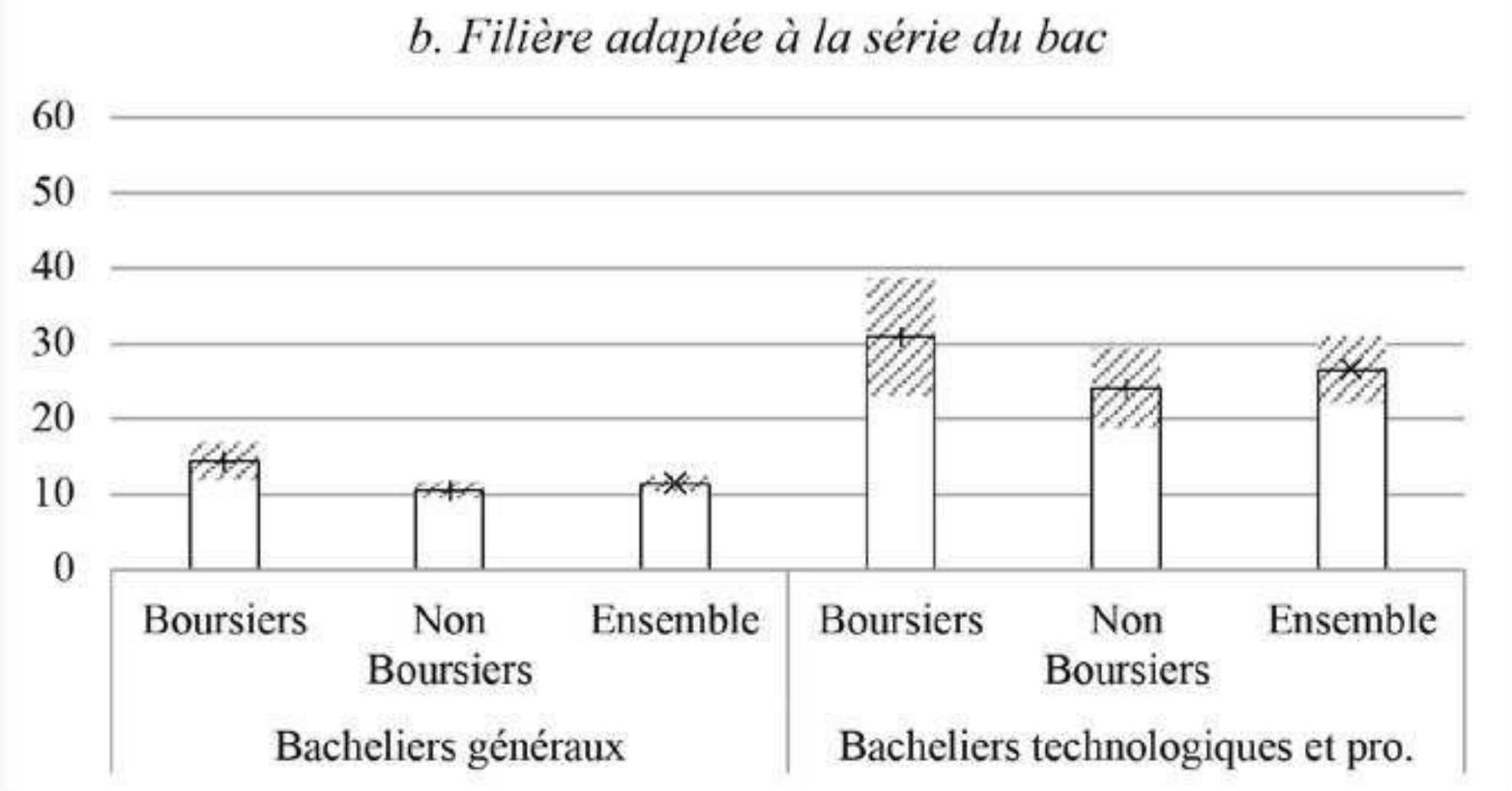

+ Khi2 de Rao-Scott : 58,6 ddl : 3 p < 0,01
× Khi2 de Rao-Scott : 48,8 ddl : 1 p < 0.01

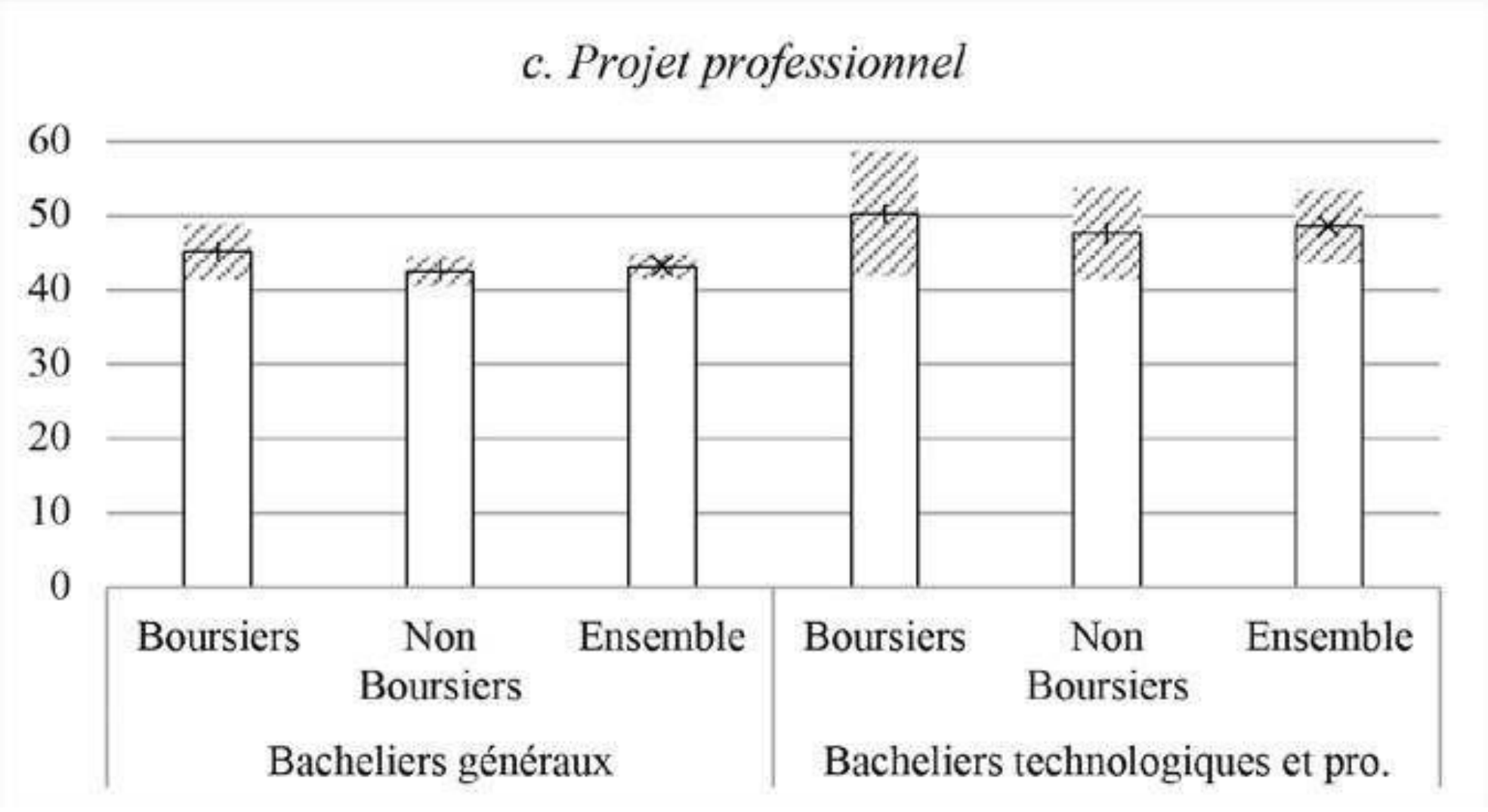

+ Khi2 de Rao-Scott : 4 ddl : 3 p > 0,1 n.s
× Khi2 de Rao-Scott : 2,8 ddl : 1 p < 0.1

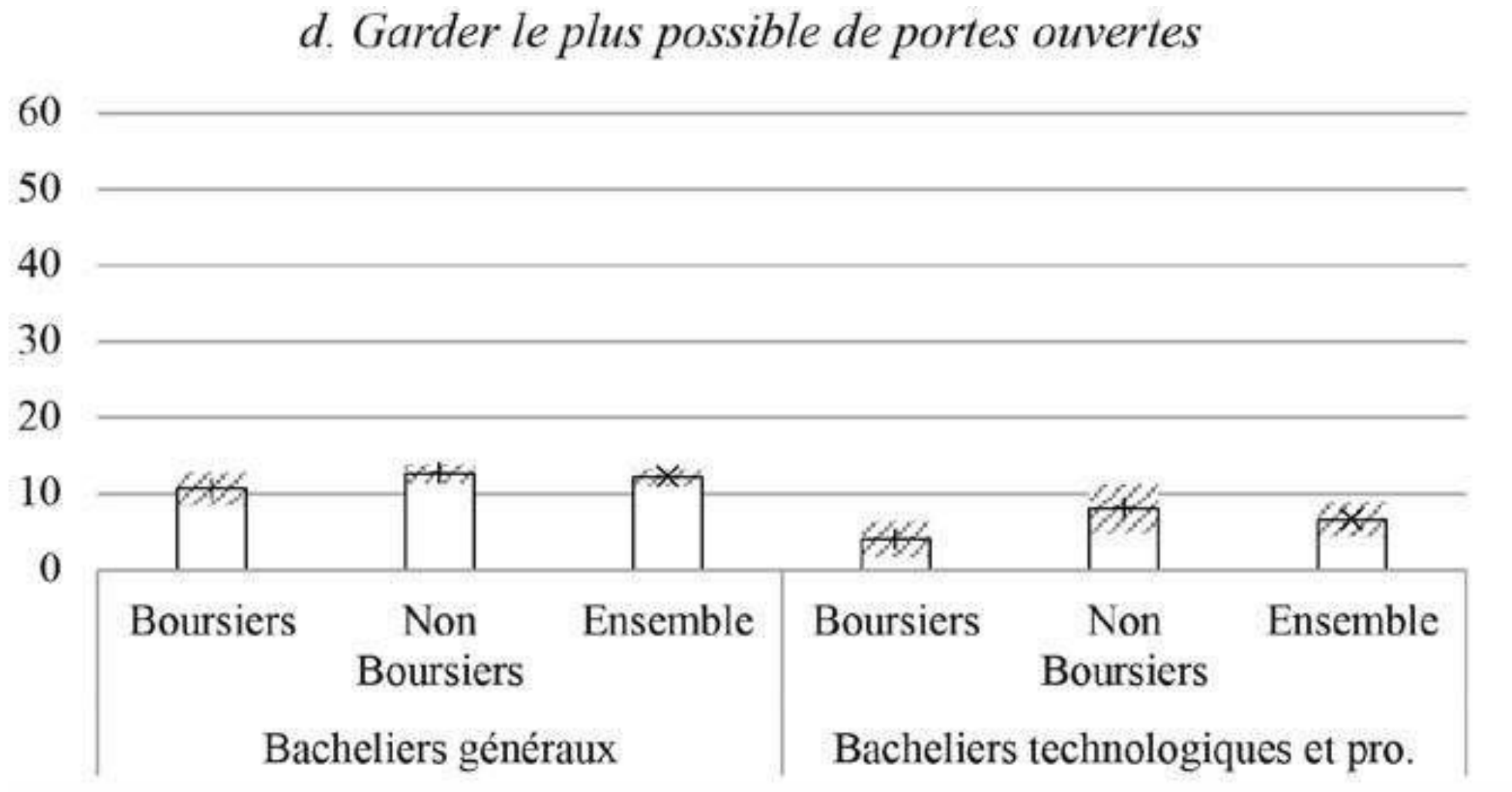

+ Khi2 de Rao-Scott : 14,2 ddl : 3 p < 0,01
× Khi2 de Rao-Scott : 9 ddl : 1 p < 0.01
Source : Enquête TSS.
Champ : Lycéens (2017-2018) admis dans l'enseignement supérieur *via* le portail Parcoursup (n =3978).
Note : dans TSS, une liste de 15 items étaient proposées aux étudiants pour chacun de leurs trois vœux favoris. Cette figure présente les items les plus fréquemment avancés et/ou qui font sens au regard de l'analyse des projets.

Ces rapports différenciés aux études supérieures conduisent les bacheliers à formuler des vœux sur des segments distincts de l'enseignement supérieur, mais ils s'expriment aussi dans la manière d'appréhender l'accès à une même formation.

*Des rapports différenciés à un même cursus : le cas d'une licence de sociologie*

L'analyse textuelle des lettres de motivation donne à voir deux oppositions principales qui sont lisibles sur les deux axes du graphique (Figure 2).

Figure 2 - L'espace du vocabulaire mobilisé dans les lettres de motivation en licence de sociologie

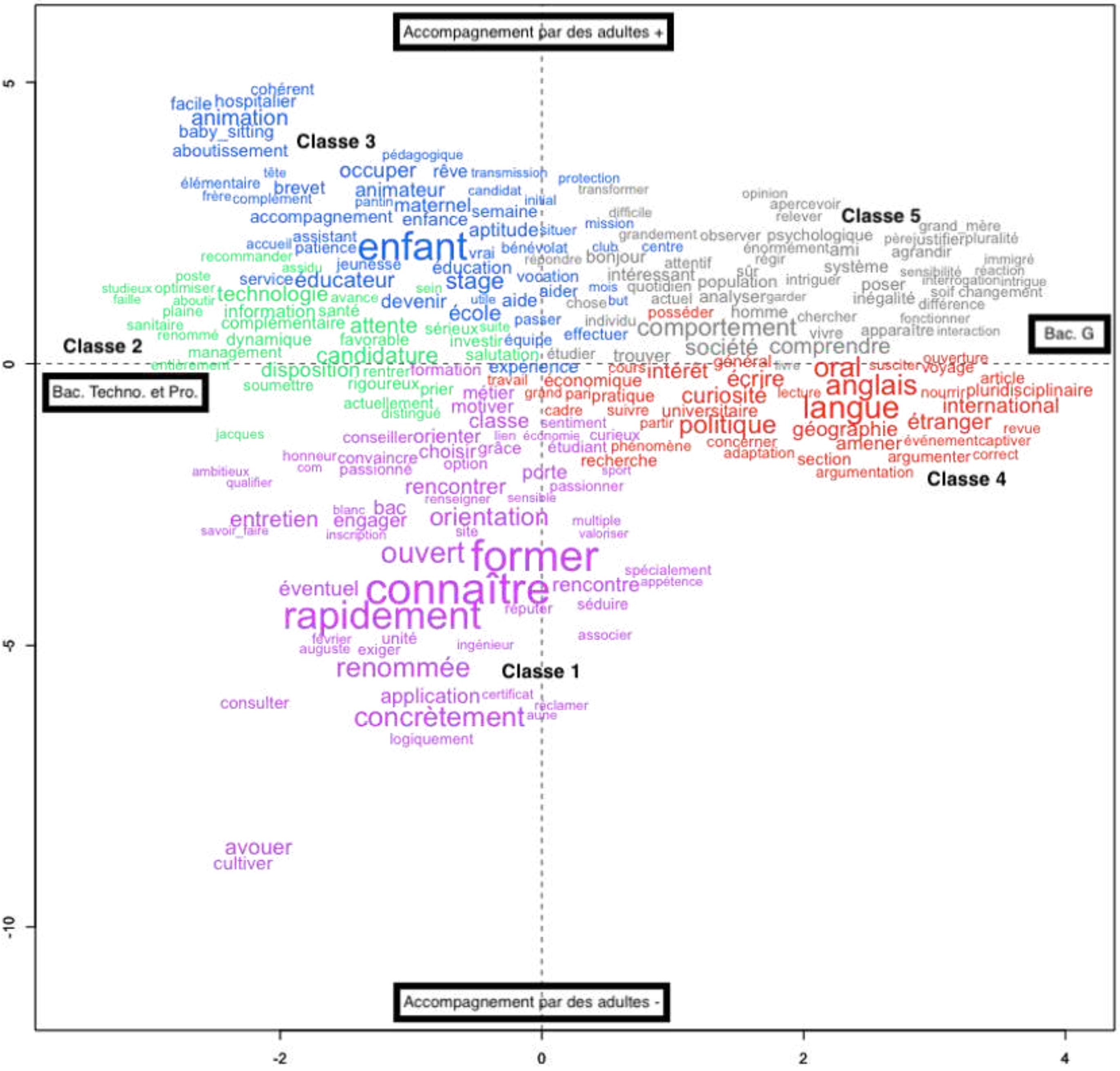

Source : Dossiers Parcoursup des candidats en licence de sociologie en 2018-2019.

Champ : Les 743 lycéens ayant candidaté dans la licence de sociologie lors de la première phase de Parcoursup en 2018.

Note : Les classes de la CHD apparaissent le plan de l'AFC dans différentes couleurs sur la version en ligne de l'article et sont identifiées par un numéro : classe 1 en bas en violet ; classe 2 à gauche en vert ; classe 3 en haut à gauche en bleu ; classe 4 à droite en bas en rouge ; et classe 5 à droite en haut en gris. La taille des mots est corrélée au chi-deux de liaison avec la classe, c'est-à-dire que plus les mots apparaissent dans une grande taille, plus ils sont spécifiques à la classe de la CHD.

Nous n'analysons que le premier plan factoriel mais les troisième et quatrième axes résument respectivement 21 % et 18 % de l'inertie.

Le 2e axe (25 % de l'inertie) exprime une opposition entre, au sud, des projets motivés « génériques », largement inspirés de lettres types accessibles en ligne et, au nord, des lettres de motivation plus personnalisées ou ajustées à la licence de sociologie. La structure cet axe est assez peu lisible avec les variables supplémentaires, bien qu'au sud de l'axe se situent principalement les élèves qui ont échoué au bac cette année-là. Une hypothèse pour expliquer sa structuration se situe dans le degré d'accompagnement (par les professeurs ou les proches) dont ont bénéficié les élèves. En effet, pour nombre de bacheliers la rédaction d'une lettre de motivation constitue une expérience nouvelle et le recours à des personnes familières de l'exercice peut s'avérer crucial. En l'absence d'un tel soutien, l'utilisation d'internet reste une alternative. La classe 1, en violet – qui regroupe 11 % des projets – s'appuie donc essentiellement sur des ressources en ligne. Les expressions « je suis déterminé à me former rapidement », « je me tiens à votre disposition pour un entretien » sont surreprésentées[12] dans cette classe car elles sont empruntées à des modèles trouvés sur internet. Il en va de même des références aux « rencontres » durant la « journée portes ouvertes » de l'établissement, d'autant que celle-ci a été finalement annulée. Les élèves qui ont échoué au baccalauréat, caractéristique de cette classe, n'ont pas été interrogés dans l'enquête TSS. Nous nous concentrerons donc sur les autres groupes dans la suite de l'article.

Le 1er axe contribue à 36 % de l'inertie. Il est structuré par une opposition entre, à gauche, les projets motivés centrés sur l'insertion professionnelle dans un métier du care et, à droite, ceux qui mobilisent plutôt du vocabulaire conceptuel de la sociologie (« Bourdieu », « inégalités », « système », « processus », « comportement » et « société ») ou qui citent des pratiques socialement valorisées (voyage, débat et parler anglais par exemple). Ce premier axe sépare très clairement les élèves selon leur série de baccalauréat opposant ainsi les candidats des voies technologiques ou professionnelles à gauche et les bacheliers généraux à droite. L'axe repose également sur une opposition genrée qui est liée au fait que les candidats en bac technologique et professionnel sont essentiellement des candidates, inscrites notamment en ST2S[13] ou en ASSP[14]. L'analyse textuelle recoupe ici les observations issues de l'enquête TSS sur les motivations à l'origine des vœux en distinguant, d'un côté, les lettres centrées sur le contenu de la discipline (dans sa dimension théorique) caractéristiques des bacheliers généraux et de l'autre, des lettres qui mentionnent un projet professionnel en adéquation avec le diplôme préparé (ST2S

---

[12] La surreprésentation dans les classes est toujours calculée par rapport à la représentation dans l'ensemble des 743 candidatures. On considère qu'il y a une surreprésentation à partir du moment où la p-value associée au chi-deux d'association avec la classe est strictement inférieure à 0,01.

[13] Sciences et Technologies de la Santé et du Social

[14] Accompagnement, Soins et Services à la Personne

ou ASSP) par leurs autrices[15].

La série du baccalauréat génère visiblement des rapports contrastés à l'enseignement supérieur qui s'expriment à l'entrée de la formation considérée dans les projets motivés. Pour rédiger ces lettres, les lycéens n'ont pas bénéficié du même accompagnement. Les pratiques des enseignants et des proches sont donc importantes pour comprendre certaines différences internes à ce groupe d'élèves dans l'expression de leurs aspirations sur Parcoursup.

## Le rôle décisif de l'École pour les bacheliers technologiques et professionnels

La réforme prévoit un accompagnement individualisé par deux professeurs principaux et des temps dédiés à la procédure d'orientation. Les élèves les moins armés scolairement et socialement sont largement dépendants de cet accompagnement pour plusieurs raisons, déjà identifiées dans la littérature mais que l'enquête TSS vient ici confirmer ou préciser.

*Internalisation de la procédure d'orientation et ciblage des meilleurs élèves*

D'abord, les élèves de classes populaires sont contraints de s'en remettre aux recommandations de leurs enseignants dans la construction de leur orientation post-bac dans la mesure où leurs proches, ne disposent pas des ressources ou du sentiment de légitimité nécessaires à leur accompagnement. Ainsi, le tableau 1 montre que les élèves les moins dotés socialement recourent moins souvent à leurs parents (Belghith et al., 2019, p. 5). En raison de la ségrégation interne à l'enseignement secondaire, ces lycéens sont nombreux en séries technologiques et professionnelles. Dans ces voies, seuls 47 % des élèves disposent d'au moins un parent diplômé de l'enseignement supérieur contre 67 % en général[16]. Les élèves qui n'ont d'autres choix que de s'appuyer sur les informations fournies par les équipes pédagogiques y sont donc davantage représentés.

[15] Notons que dans l'enquête TSS les femmes des voies technologiques et professionnelles évoquent plus souvent le projet professionnel comme une source de motivation que les hommes (pour 55 % d'entre elles contre 42 % des garçons) et qu'il n'existe pas de telles différences genrées au sein des séries générales.

[16] Notons aussi que les voies technologiques et professionnelles accueillent davantage d'enfants d'immigrés, respectivement 34 et 39 %, contre 22 % en générales. Ces parents constituent une population potentiellement moins familière encore du système éducatif français.

*Tableau 1 - Recours aux enseignants ou aux parents comme source d'information (en %)*

| | Recours aux enseignants | | Recours aux parents | |
|---|---|---|---|---|
| | *Estimation* | *Intervalle de confiance à 90 %* | *Estimation* | *Intervalle de confiance à 90 %* |
| *Statut de boursier* | *Khi2 : 2,3 ddl : 1 p > 0,1 n.s* | | *Khi2 : 23,9 ddl : 1 p < 0,01* | |
| Boursiers | 37 | [34-41] | 14 | [12-17] |
| Non Boursiers | 33 | [32-35] | 24 | [23-26] |
| *Niveau d'études le plus élevé des parents* | *Khi2 : 4,2 ddl : 1 p < 0,05* | | *Khi2 : 43,2 ddl : 1 p < 0,01* | |
| Diplômé équivalent au bac ou d'un niveau inférieur | 37 | [34-40] | 14 | [12-16] |
| Diplômé du supérieur | 32 | [30-34] | 27 | [25-29] |

Source : Enquête TSS.
Champ : Lycéens (2017-2018) admis dans l'enseignement supérieur via le portail Parcoursup (n =3978).

Par ailleurs, l'accompagnement à la rédaction des projets motivés est davantage pris en charge par les professeurs principaux au sein des filières technologiques et professionnelles en comparaison aux voies générales : 45 % des bacheliers technologiques et professionnels assurent avoir été aidés par leurs enseignants dans la composition de leurs lettres contre 34 % des bacheliers généraux (Figure 3)[17]. Ce résultat atteste d'une internalisation plus forte de la procédure au sein des voies technologiques et professionnelles en comparaison aux filières générales. Lemêtre et Orange (2017, p. 59) l'observaient déjà au moment de saisir les vœux sur l'ancien portail APB.

[17] Nous observons systématiquement les mêmes dynamiques pour l'accompagnement à la rédaction des CV que nous ne commentons pas ici.

Figure 3- Aide des enseignants dans la rédaction des projets motivés selon le type de baccalauréat et le statut de boursier (en %)

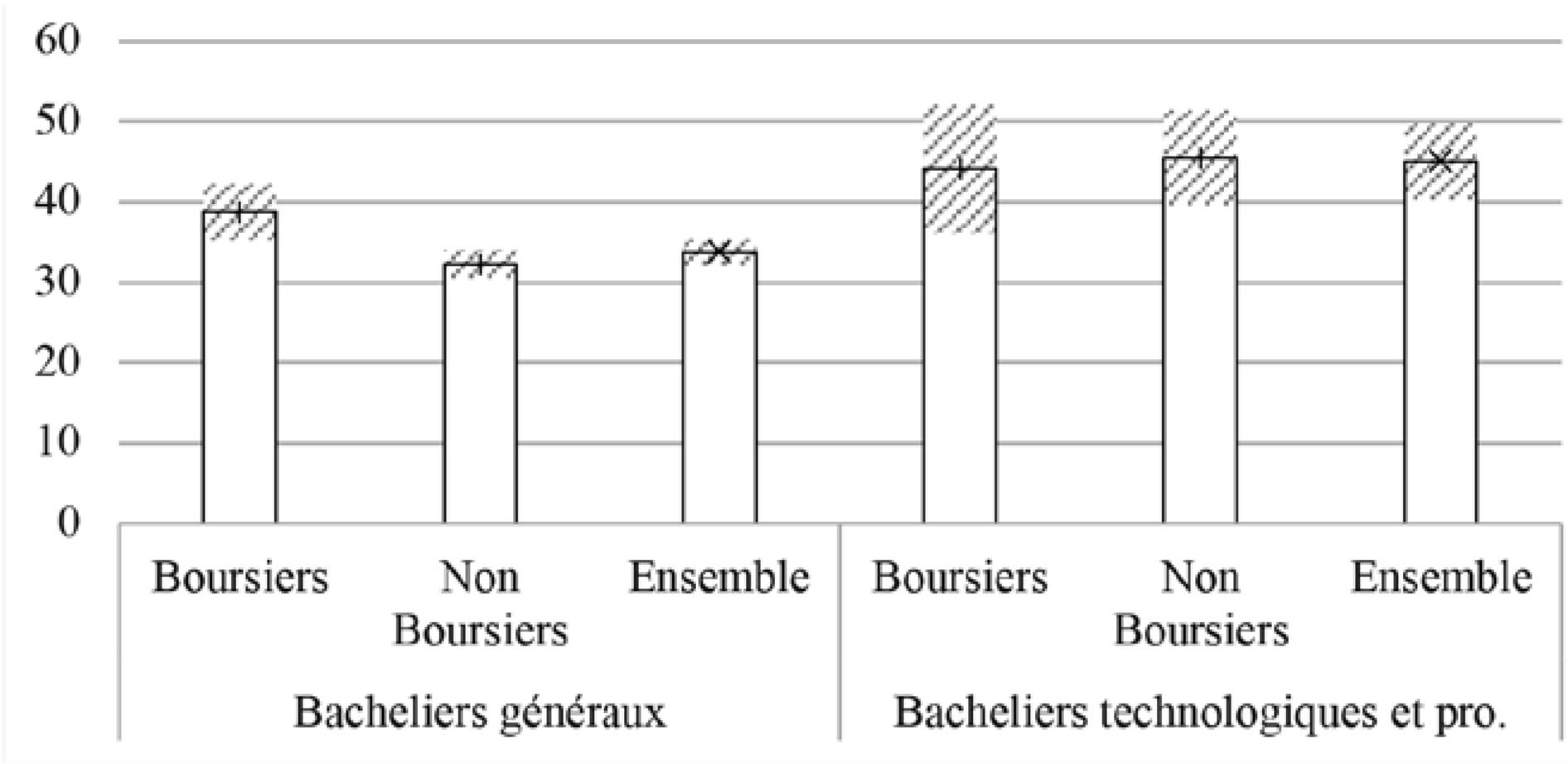

+ Khi2 de Rao-Scott : 20,8 ddl : 3 p < 0,01
× Khi2 de Rao-Scott : 14,7 ddl : 1 p < 0,01 Source : Enquête TSS.
Champ : Lycéens (2017-2018) admis dans l'enseignement supérieur via le portail Parcoursup (n =3978).

De fait, l'École est centrale dans l'encadrement des projets de poursuite d'études des bacheliers technologiques et professionnels et elle l'est davantage encore pour les lycéens marqués ou identifiés comme de bons élèves. Comme décrit par Truong (2013) dans le cas de bacheliers généraux, les enseignants semblent concentrer leurs efforts sur certains élèves disposant d'un bon dossier et auprès desquels ils ont le sentiment d'être utile. Plusieurs résultats de l'enquête TSS confortent cette idée, mais au sein des terminales technologiques et professionnelles uniquement. On y observe en effet une forte hétérogénéité des pratiques d'encadrement selon le niveau scolaire des élèves[18] : quand 65 % des bacheliers technologiques et professionnels disposant des meilleures mentions disent avoir échangé personnellement avec leurs professeurs principaux au sujet de Parcoursup, seuls 36 % de leurs homologues sans mention partagent le même souvenir (Figure 4). En revanche, on n'identifie pas de différence notable selon les résultats des élèves en terminale générale : 53 % des bacheliers généraux affirment avoir échangé personnellement avec leurs professeurs principaux, peu importe leur mention. Dans ce contexte, les meilleurs bacheliers technologiques et professionnels sont plus nombreux que les autres à considérer que leurs enseignants les ont aidés à clarifier leur perspective (Figure 5).

[18] Le niveau scolaire des élèves est mesuré ici par la mention obtenue au baccalauréat.

*Figure 4 - Échange(s) personnalisé(s) avec les professeurs principaux concernant Parcoursup (en %)*

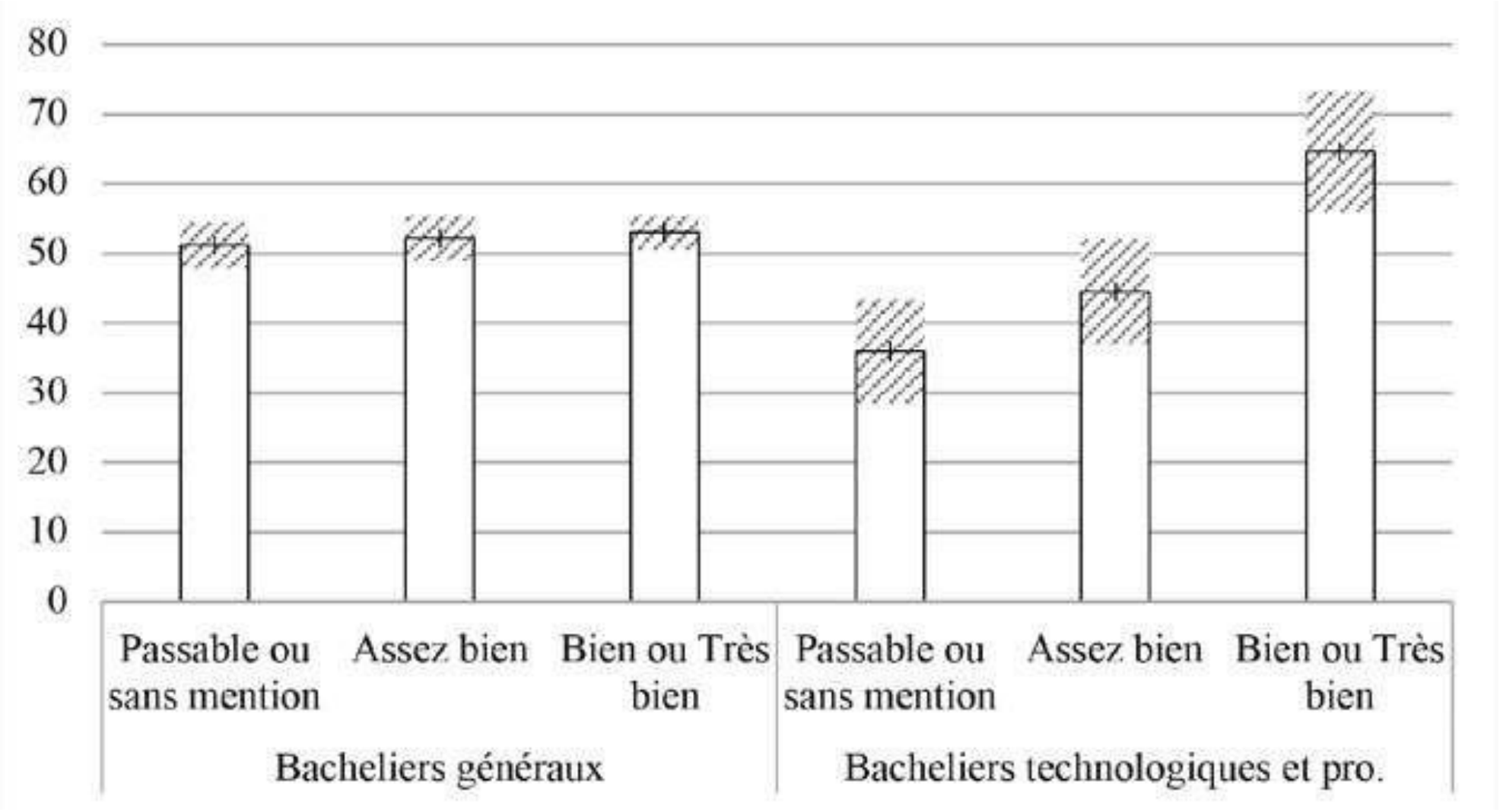

+ Khi2 de Rao-Scott : 24,4 ddl : 5 p < 0,01 Source : Enquête TSS.
Champ : Lycéens (2017-2018) admis dans l'enseignement supérieur via le portail Parcoursup (n =3978).

*Figure 5 - Aide des enseignants dans la clarification de l'orientation selon le type de baccalauréat et le niveau scolaire (en %)*

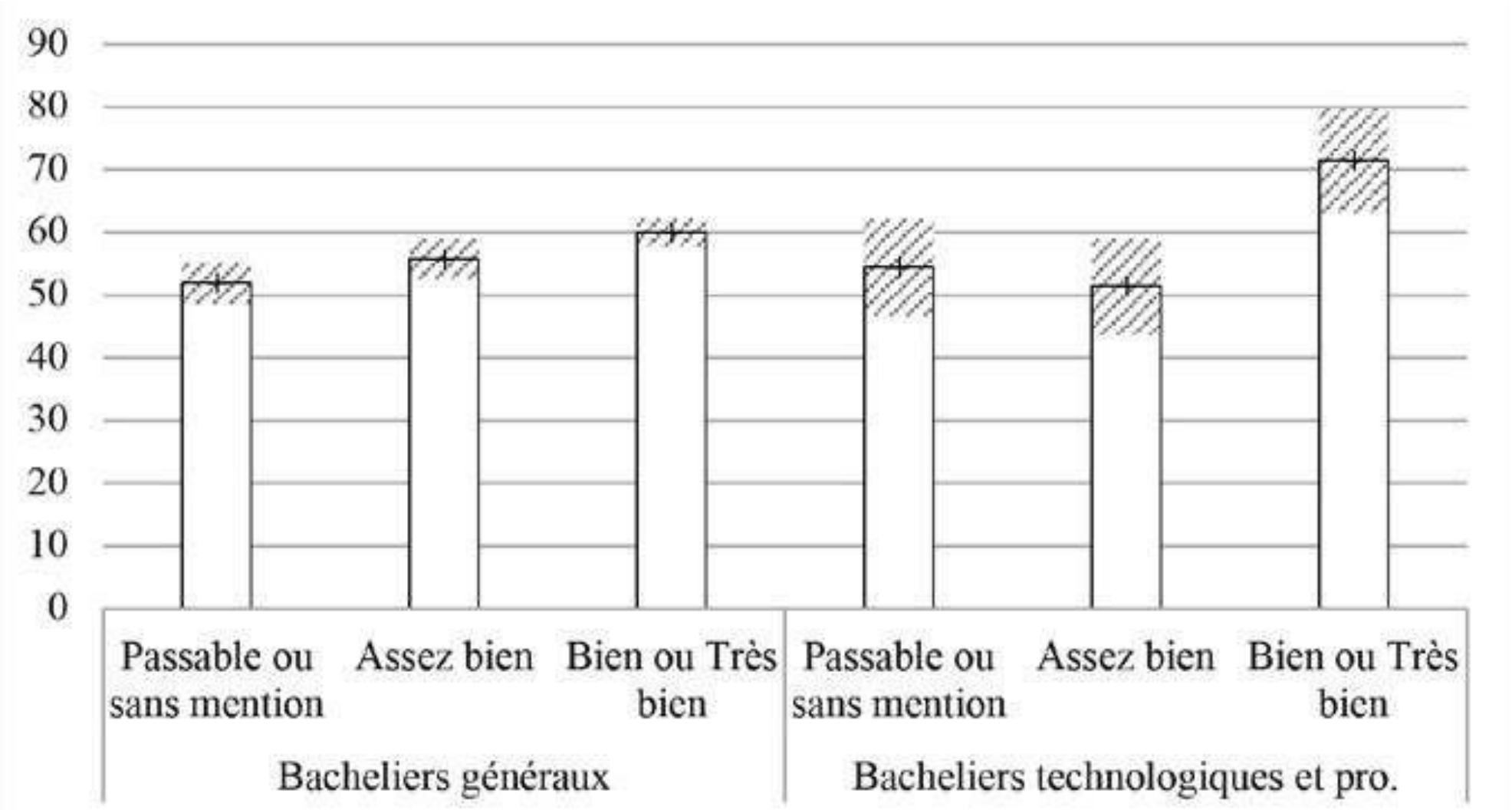

+ Khi2 de Rao-Scott : 16,7 ddl : 5 p < 0,01 Source : Enquête TSS.
Champ : Lycéens (2017-2018) admis dans l'enseignement supérieur via le portail Parcoursup (n =3978).

La figure 6 montre la même tendance concernant la rédaction des projets motivés avec un investissement plus marqué des enseignants auprès des élèves les mieux évalués aux bacs technologiques et professionnels uniquement. Alors qu'en voie générale le niveau scolaire ne détermine pas ou peu l'aide des professeurs référents, dans les sections technologiques et professionnelles les meilleurs lycéens déclarent un accompagnement renforcé de la part de leur enseignant comparativement aux autres élèves. Une explication possible (mais non suffisante) de ce phénomène tient à l'éthos professionnel des professeurs en lycée professionnel[19]. Une autre renvoie à la position singulière des bons élèves des séries technologiques et professionnels dans le système éducatif français : ils sont désavantagés dans la compétition scolaire en raison de leur série au baccalauréat tout en disposant de résultats scolaires leur permettant de concurrencer les lycéens des filières générales à l'entrée de certaines formations (et/ou pour les élèves des voies professionnelles, les bacheliers technologiques[20]). Autrement dit, dans un contexte où les ressources allouées pour accompagner les élèves sont limitées, les professeurs principaux semblent cibler certains lycéens auprès desquels ils considèrent leur aide décisive. Cet investissement ciblé a pour effet d'influencer les vœux de poursuite d'études comme Truong (2013) le décrit précisément pour les élèves de classe générale qui sont repérés et encouragés à postuler dans les filières sélectives de l'enseignement supérieur. Ainsi, dans l'enquête TSS, 41 % des bacheliers technologiques et professionnels félicités par une mention Très bien ou Bien désignent un IUT comme vœu favori, contre 17 % de leurs homologues sans mention.

*Figure 6 - Aide des enseignants dans la rédaction des projets motivés selon le type de baccalauréat et le niveau scolaire (en %)*

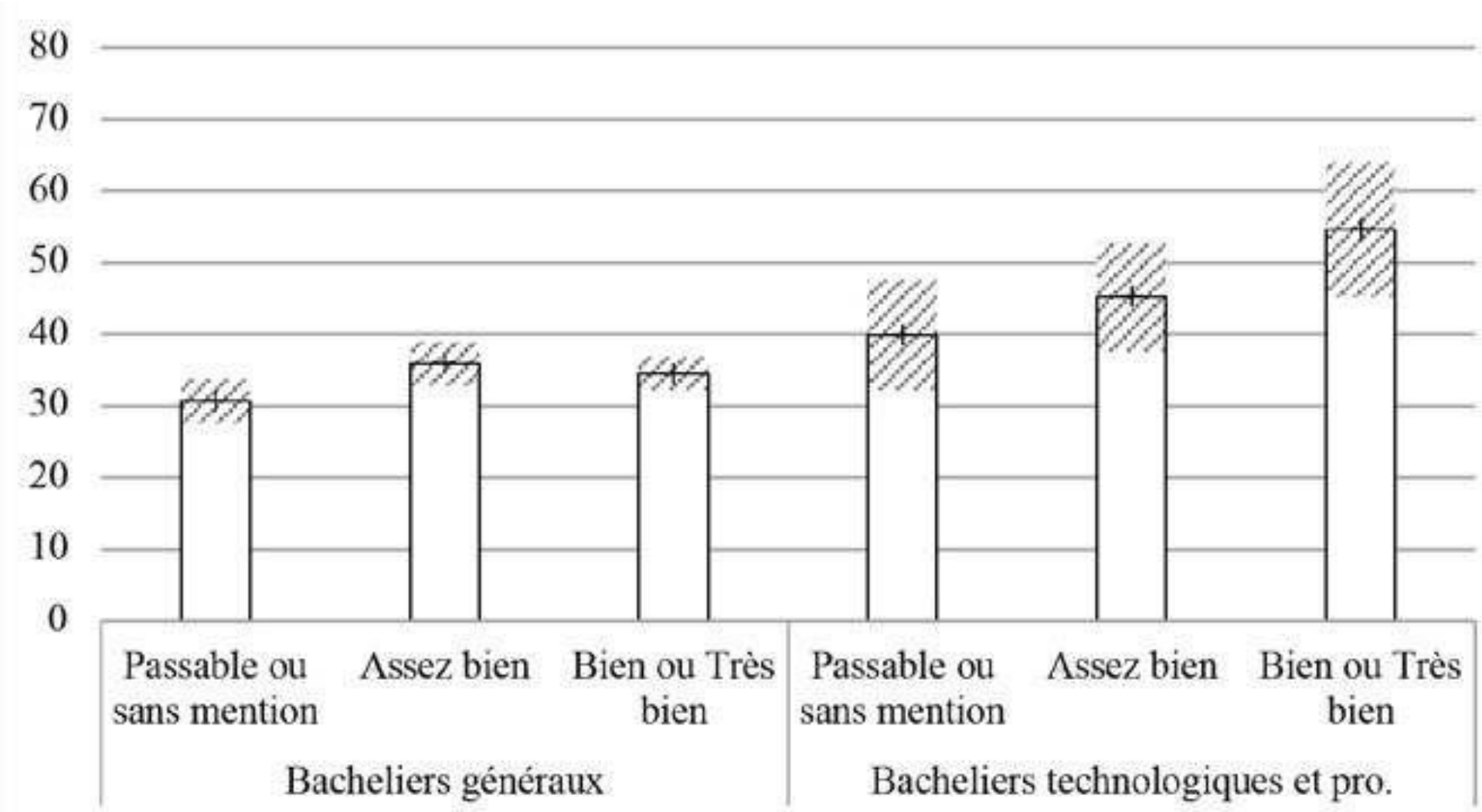

+ Khi2 de Rao-Scott : 25,3 ddl : 5 p < 0,01

Cet accompagnement spécifique peut donc amener les bons élèves des séries technologiques et professionnelles à envisager des formations différentes de leurs camarades qui ne font pas l'objet d'un repérage professoral. Symétriquement, lorsqu'ils candidatent aux mêmes formations, l'encadrement par les enseignants les conduit à exprimer différemment leurs aspirations scolaires dans les lettres qu'ils rédigent.

[19] Cette interprétation est non-suffisante car les tendances décrites s'observent dans les voies professionnelles comme technologiques, même si les effectifs restreints de ces populations prises isolément ne nous permettent pas d'estimer la tendance avec suffisamment de précision.

[20] Peu importe ici à quels élèves les équipes pédagogiques ou les élèves se comparent, c'est une position relative perçue comme moins favorable à l'entrée de certaines formations qui structurait les pratiques.

*La marque des enseignants sur les projets motivés*

L'analyse des lettres de motivation en licence de sociologie distingue deux sortes de projets ou classes parmi les élèves de bacs technologiques ou professionnels – en vert et bleu à gauche du graphique – en partie liée au capital scolaire des élèves (Figure 2).

La classe 2, en vert, rassemble le plus de candidats (34 %). Les élèves qui constituent ce groupe sont plus souvent des candidats préparant un baccalauréat technologique et l'ayant finalement obtenu sans mention[21]. Il semble que ce groupe correspond à des élèves qui ne bénéficient ni des ressources familiales ni d'un accompagnement personnalisé par les enseignants dans la construction et l'expression de leurs vœux d'orientation post-bac.

Dans le contenu des lettres, les élèves de ce groupe se caractérisent principalement par l'usage de formulations relativement neutres avec l'utilisation de tournures de politesse : « Dans l'attente d'une réponse favorable de votre part, je vous prie d'agréer mes salutations respectueuses ». Par ailleurs, les lettres sont souvent structurées autour de l'utilisation d'adjectifs qualificatifs mobilisés par les élèves pour exprimer certaines qualités communément valorisées (« rigoureux », « patient », « studieux », « dynamique », « travailleur »).

La plus grande importance accordée par les séries technologiques et professionnelles à l'adéquation de la formation dans l'enseignement supérieur à la filière de leur baccalauréat (voir ci-dessus) se retrouve dans les lettres de motivation. Les élèves mettent en lien leur parcours scolaire antérieur avec leur candidature (« Actuellement en terminale ST2S » ou « Actuellement inscrit en bac Sciences et Technologie du Management et de la Gestion »). Les caractéristiques des élèves de cette classe semblent proches du groupe à l'orientation contrariée dans les cursus de sociologie identifié à l'aide des vœux enregistrés sur APB par Rossignol-Brunet : les « plus fragiles scolairement, à savoir celles et ceux non titulaires d'un baccalauréat général et l'ayant obtenu sans mention » (Rossignol-Brunet, 2021, p. 6). Cela expliquerait aussi les caractéristiques assez génériques de ces lettres de motivation : les élèves de ce groupe ne souhaiteraient pas prioritairement intégrer une licence de sociologie.

À proximité de cette classe, en bleu, et en haut à gauche du graphique, se retrouvent les projets motivés appartenant à la troisième classe (qui regroupe 14 % des candidats). Ces projets ont été plus souvent rédigés par des femmes et par des élèves qui ont obtenu un baccalauréat professionnel avec une mention Bien ou un baccalauréat technologique avec une mention Très Bien. Ce sont donc plus généralement les bonnes élèves des filières les moins valorisées de la hiérarchie scolaire. Il est probable que ce groupe corresponde aux élèves identifiées plus haut comme les plus soutenues par leurs enseignants puisqu'avec les meilleurs résultats.

[21] Plus précisément : sans mention Bien ou Très Bien.

Comme pour le groupe précédent, les lettres de cette classe insistent particulièrement sur la continuité entre la trajectoire scolaire antérieure et la licence de sociologie. Cela passe notamment par l'évocation de l'acronyme « ASSP » par les élèves qui ont suivi ce cursus ou par l'évocation des différents « stages » réalisés pendant leur formation au lycée. D'autre part, les élèves valorisent des activités extrascolaires en lien avec leur formation initiale (dans le domaine sanitaire et social). Le groupe se caractérise ainsi par de fréquentes évocations d'expériences antérieures de travail éducatif soit dans un cadre institutionnalisé (« BAFA ») ou semi-institutionnalisé (« babysitting »), soit dans le cadre familial (expériences de garde des « frères » et sœurs). Sont également mentionnées des dispositions genrées liées au care : les élèves écrivent être doués pour « écouter », « aider » ou « accompagner ». Les lettres mentionnent aussi des projets professionnels en adéquation avec le baccalauréat de leurs autrices et leur vœu en sociologie. Le souhait d'accéder à un statut de la fonction publique se retrouve, soit lorsque les candidats précisent qu'ils espèrent intégrer la « PJJ » (Protection Judiciaire de la Jeunesse) soit en évoquant des professions de travail éducatif (« animation », ou enseignant en « maternelle », « éducatrice[22] » ou « éducatrice spécialisée »). Ce groupe semble ainsi correspondre aux étudiants qui prépareront un concours pendant leurs années d'études, l'un des six profils d'étudiants en sociologie identifiés par Millet à la fin des années 1990 (Millet, 2000) ; ces derniers envisagent la licence de sociologie comme une passerelle pour des concours ou des formations sélectives « en lien avec des problématiques sociales, sanitaires et/ ou éducatives » (Rossignol-Brunet, 2021, p. 7).

Les autrices de ces lettres ont répondu aux consignes officielles du projet motivé en mentionnant des stages, des activités extrascolaires et des qualités (ici particulièrement féminines) qu'elles estiment valorisables dans le cadre d'un projet professionnel (dans le domaine sanitaire et social). Elles s'efforcent de mettre en cohérence leur formation dans le secondaire avec leur vœu en sociologie. Elles semblent donc avoir intégré certaines attentes (au moins) de l'institution concernant leur orientation scolaire et notamment celles véhiculées par leurs enseignants.

Le projet motivé de Malorie est l'un des plus caractéristiques de cette classe. Malorie préparait en 2017-2018 un baccalauréat professionnel ASSP et était boursière du secondaire. Elle a obtenu son baccalauréat avec la mention Bien et l'analyse de son dossier scolaire traduit un investissement professoral particulier. En effet, sur sa fiche Parcoursup, le conseil de classe a fait figurer un commentaire général facultatif qui indique qu'elle « cadre avec le monde du social ». D'autre part, les appréciations sur ses bulletins de lycée traduisent une satisfaction de la part de ses professeurs (« Bon travail », « Très bien », « Élève agréable et sérieuse. Très bon semestre. ») et elle a obtenu les félicitations du conseil de classe lors du second semestre de terminale. Dans sa lettre, elle présente son souhait d'intégrer la formation en le mettant en cohérence avec ses dispositions genrées (et qu'elle nomme des « valeurs ») : « Le secteur du social représente un milieu ou l'accompagnement, le soutien, l'entraide sont omniprésent et font partie intégrante de mes valeurs, d'où mon intérêt pour votre License. (Sic) ». Par ailleurs, Malorie lie son souhait d'intégrer la licence avec son parcours scolaire (« J'aimerais intégrer votre établissement, cela serait pour moi l'aboutissement de trois années d'études ») et les stages qu'elle a réalisés (« J'ai eu l'occasion durant mes années scolaires d'effectuer de nombreux stages dans des EHPAD, ainsi que dans le milieu de la petite enfance. »). Enfin, elle évoque également ses souhaits d'insertion professionnelle dans le « secteur du social ». L'accompagnement professoral dont Malorie semble avoir bénéficié la conduit ainsi à envisager ses études dans la continuité directe de son baccalauréat professionnel.

[22] Le logiciel Iramuteq procède à une lemmatisation des termes, c'est-à-dire que les noms, les verbes et les adjectifs sont comptabilisés dans une forme unique (le masculin singulier pour les noms et les adjectifs et l'infinitif pour les verbes). Ainsi le terme « éducatrice » revient plus souvent dans sa forme féminine, du fait des caractéristiques des candidats de ce groupe, qui sont plus souvent des candidates mais c'est le terme « éducateur », dans la forme masculine, qui apparaît dans le graphique.

La lettre de Malorie, comme celles des autres candidates de son groupe, montre donc une intériorisation des attentes de l'institution scolaire concernant son orientation ou a minima la mobilisation de ressources ou de recommandations transmises par ses professeurs. En effet, sa lettre respecte de nombreuses consignes officielles du projet motivé, contrairement à celles, plus évasives, des élèves de la deuxième classe de la CHD qui se caractérisent par de moins bons résultats scolaires et, d'après l'enquête TSS, par un moindre accompagnement des enseignants. Ainsi, les bonnes élèves des séries technologiques et professionnelles sont peut-être davantage disposées à suivre les recommandations, mais elles sont aussi davantage accompagnées par l'École pour le faire. Si le poids de l'École est notable ici – par l'encadrement différencié des élèves, ainsi que sa marque sur les projets motivés – la situation est différente pour les bacheliers généraux.

## L'importance des ressources familiales dans l'orientation des bacheliers généraux

Les bacheliers généraux appliquent d'une autre manière encore les consignes relatives au projet motivé en éludant la question de leurs plans professionnels, mais en attestant d'une forme de familiarité avec le monde universitaire. Comme pour les bacheliers technologiques et professionnels, le corpus composé de leurs lettres se scinde en deux groupes selon le degré d'adaptation à la formation visée. Mais dans le cas des lycéens généraux, il apparaît que les contrastes internes tiennent davantage à l'origine sociale des élèves qu'à leur capital scolaire. Les modalités d'accompagnement en terminale générale, observées dans l'enquête TSS, semblent expliquer l'importance des ressources familiales dans la structuration du corpus dans la mesure où une partie du travail d'orientation est confiée aux parents.

*Déléguer une partie du travail d'orientation aux familles*

Les élèves non-boursiers sont plus nombreux que les autres à se reposer sur l'aide exclusive de leur famille pour rédiger leurs lettres. Plus précisément, 44 % des élèves non-boursiers déclarent une telle situation, contre 31 % des élèves boursiers[23]. A contrario, l'exposé du fonctionnement technique du portail reste principalement à la charge des enseignants et cela indépendamment de l'origine sociale des élèves (Tableau 2)[24].

*Tableau 2 - Implication conjuguée de la famille et des professeurs à deux étapes différentes de la procédure selon le statut de boursier des élèves (en %)*

| | Rédaction du projet motivé | | | | Présentation du fonctionnement de Parcoursup | | | |
|---|---|---|---|---|---|---|---|---|
| | Boursiers | | Non Boursiers | | Boursiers | | Non Boursiers | |
| | *Khi2 : 65,3 ddl : 3* p *< 0,01* | | | | *Khi2 : 10,5 ddl : 3* p *< 0,05* | | | |
| Famille et Enseignants | 22 | [19-25] | 26 | [24-28] | 10 | [8-12] | 14 | [13-16] |
| Famille uniquement | 31 | [28-34] | 44 | [42-45] | 2 | [1-3] | 4 | [3-4] |
| Enseignants uniquement | 19 | [16-21] | 9 | [8-10] | 69 | [66-73] | 63 | [62-65] |
| Aucune aide | 29 | [26-32] | 22 | [20-23] | 18 | [16-21] | 19 | [17-20] |

Source : Enquête TSS.
Champ : Lycéens (2017-2018) admis dans l'enseignement supérieur via le portail Parcoursup (n =3978).

Le soutien de l'entourage apparaît particulièrement décisif dans la rédaction des lettres de motivation et notamment pour les élèves les plus dotés. Ces derniers étant structurellement plus nombreux en série générale, cela affecte les pratiques des enseignants exerçant dans ces voies. Ces derniers semblent volontiers confier l'écriture du projet motivé aux seuls élèves et à leur famille. En effet, 66 % des bacheliers généraux déclarent avoir effectué ce travail sans l'intervention de leurs professeurs, contre 55 % des bacheliers technologiques et professionnels. Les lycéens généraux ne sont pas esseulés pour autant puisqu'ils sont massivement assistés de leurs proches, pour 65 % d'entre eux (cela représente 11 points de plus que les élèves technologiques et professionnels dans la même configuration). Mais si le public des séries générales se compose d'élèves relativement dotés, il compte aussi des élèves boursiers (pour 23 %). Parmi eux, 43 % ne peuvent s'appuyer sur leurs familles en complément ou à la place de celui de leurs enseignants et 28 % sont démunis de ces deux formes de soutien. Comparativement, les lycéens généraux plus favorisés sont respectivement 29 % et 22 % dans ce cas (Tableau 3).

[23] Si les élèves boursiers peuvent compter sur l'appui de leurs professeurs, 29 % restent complétement démunis face à l'exercice du projet motivé. Seuls 22 % des élèves non-boursiers se trouvent dans une telle situation.

[24] Les familles privilégiées viennent compléter de manière très sensible seulement les informations fournies au sein des lycées. On observe le même phénomène pour la présentation générale des différents choix d'orientation.

*Tableau 3 - Implication conjuguée de la famille et des professeurs dans la rédaction des projets motivés selon le statut de boursier et le type de baccalauréat des élèves (en %)*

| | Bacheliers généraux | | | | Bacheliers technologiques et pro. | | | | |
|---|---|---|---|---|---|---|---|---|---|
| | Boursiers | | Non Boursiers | | Boursiers | | Non Boursiers | | |
| | *Khi2 : 46 ddl : 3* p < *0,01* | | | | *Khi2 : 14,2 ddl : 3* p < *0,01* | | | | |
| Famille et Enseignants | 24 | [21-27] | 25 | [23-27] | 17 | [11-22] | 31 | | [25-36] |
| Famille uniquement | 33 | [30-37] | 46 | [44-48] | 25 | [18-32] | 33 | | [27-38] |
| Enseignants uniquement | 15 | [12-18] | 7 | [6-8] | 27 | [20-35] | 15 | | [10-20] |
| Aucune aide | 28 | [25-31] | 22 | [20-23] | 31 | [24-38] | 22 | | [17-27] |

Source : Enquête TSS.
Champ : Lycéens (2017-2018) admis dans l'enseignement supérieur via le portail Parcoursup (n =3978).

Dans les séries générales, l'accompagnement professoral est donc plus faible de manière générale et, on l'a vu, ne tient pas aux résultats scolaires des élèves. La procédure d'orientation est davantage externalisée au sens où elle est prise en charge par la famille, notamment par les familles les plus privilégiées. Dans ce contexte, les bacheliers généraux présentent leurs proches comme déterminant dans la clarification des choix d'orientation, pour 62 % d'entre eux contre 49 % des lycéens technologiques et professionnels. Au moment de préciser les choix, de rédiger le projet de motivation, les élèves des séries générales sont ainsi renvoyés aux ressources culturelles et sociales dont ils disposent en dehors de l'École. Les inégalités entre classes sociales jouent donc plus nettement dans ces filières générales et ces inégalités sont visibles dans le vocabulaire mobilisé dans les lettres de motivation.

*La marque des dispositions familiales sur les projets motivés*

L'analyse des lettres de motivation en licence de sociologie fait apparaître deux classes, parmi les élèves de terminale générale – en rouge et gris à droite du graphique (Figure 2) – qui se distinguent entre elles par le statut de boursier des élèves. Dans la classe 4 en rouge, les élèves boursiers du supérieur sont surreprésentés tandis qu'au contraire, les élèves de la dernière classe, se caractérisent notamment par une sous-représentation des élèves boursiers du supérieur comme du secondaire.
Le groupe 4 compte nombre de bacheliers avec mention et de lycéens généraux boursiers du supérieur. Cette classe est donc susceptible de rassembler des élèves dont les familles disposent de peu de ressources, mais qui n'ont pas été accompagnés par leurs enseignants. Leurs projets font état de dispositions et de pratiques socialement et scolairement valorisées mais qui ne sont pas adaptées à la licence où ils candidatent. Les élèves de ce groupe mentionnent des pratiques internationales (« étranger », « voyage », « anglais », etc.) ou évoquent des dispositions d'expression scolaires (« écrit » et « oral »). Quelques références au milieu universitaire sont également mobilisées (« universitaire », « scientifique », « recherche ») ainsi qu'à l'ensemble des disciplines des sciences sociales (« géographie », « histoire », « sciences politiques », « anthropologie », « économie » et même « pluridisciplinaire »). Les élèves évoquent aussi des pratiques culturelles valorisées scolairement comme regarder des « reportages », écrire/lire des « articles » de presse ou de blog. Les éléments cités renvoient donc à des intérêts et des pratiques légitimes du point de vue de l'institution scolaire. L'écriture des lettres de motivation s'apparente ainsi à un exercice scolaire de mises en exergue de compétences scolairement rentables. Les élèves n'évoquent pas d'éléments de leurs trajectoires personnelles, ni de parcours professionnels dans lesquels s'intégrerait la licence de sociologie. En revanche, on retrouve dans ces lettres l'intérêt plus nettement exprimé par les bacheliers généraux pour le contenu des études pour justifier leurs choix de formation.

Les élèves de la dernière classe, en gris, représentent 16 % des candidats de la licence. Ce groupe se distingue des autres par une surreprésentation des lycéens généraux et plus spécifiquement non-boursiers. La classe se caractérise également par une plus grosse proportion de bacheliers mention Assez Bien (mais pas Bien ou Très Bien) et d'hommes. Les élèves semblent définis davantage par leur origine sociale favorisée que par leur réussite scolaire et nous supposons, à partir des données de l'enquête TSS, qu'il s'agit des élèves de classes supérieures particulièrement accompagnés par leurs proches dans la rédaction de leurs lettres. Les projets motivés sont ainsi particulièrement personnalisés pour la sociologie, par contraste avec la classe précédente. Ils contiennent des termes clés ou des concepts de sciences sociales que les élèves mobilisent pour décrire avec précision la formation à laquelle ils et elles candidatent (« comprendre » les « comportements », de l'« homme »/des « individus »/ des « gens » en « société »), mais aussi des descriptions des méthodes d'enquête (« observation »). Ou encore, elles contiennent des références légitimes à des chercheurs ou chercheuses en sociologie (« Bourdieu » ou « Isabelle Clair »). Les lettres mobilisent aussi le champ lexical de la curiosité pour justifier l'intérêt personnel de l'élève pour la discipline (« soif de connaissance », être « intrigué »). Enfin, les projets font également mention des parcours personnels des élèves, lesquels sont mobilisés de façon à motiver leur intérêt pour les sciences sociales. Quelques-uns reviennent par exemple sur le statut d'« immigré » de personnes de leur entourage ou sur les parcours de transfuge de classe de leurs parents.

Le projet motivé de Lisa reprend les codes des lettres de motivation de cette classe. Ses parents sont membres des classes supérieures très dotées en capital culturel et sont tous deux familiers de l'enseignement supérieur. Après un an ou deux ans à l'université, le père de Lisa a rejoint l'ENS Louis Lumière, grande école publique dédiée aux métiers du cinéma. La mère de Lisa a suivi un cursus universitaire au Brésil dont elle est originaire. Lors d'un entretien avec Lisa, celle-ci confie que la production de sa lettre de motivation Parcoursup a fait l'objet d'un travail familial collectif. Il a porté aussi bien sur la forme (orthographe et syntaxe) que sur le fond (les arguments à avancer) et principalement avec son père, du fait de sa plus grande proximité avec le système scolaire français. Celui-ci l'a notamment convaincue de faire figurer dans sa lettre des éléments propres à sa trajectoire et valorisables auprès de la commission des vœux : « un exemple c'est que, moi, j'ai énormément de mal à dire que je suis franco- brésilienne et je le mets rarement. [...] Mon père [me] dit : « mais tu es complétement débile de ne pas mettre ça en avant alors que c'est hyper important ». Ce genre de truc comme ça où, moi, je suis : « oui, bon ok, c'est... ». [...] J'ai un peu l'impression de me la péter si je mets : « oui, je suis franco-brésilienne, c'est super. Je connais super bien... », enfin, non ! [...] Maintenant, je le mets parce que mon père m'a assez dit qu'il fallait le mettre. » Sur ces conseils, Lisa indique donc dans sa lettre : « ma mère étant brésilienne et mon père français, j'ai eu la chance d'avoir deux cultures et cette diversité d'environnements a aiguisé ma curiosité et mon sens de l'observation. Ceci m'a donné envie d'étudier les phénomènes de société ». Au contraire, Lisa ne se souvient pas avoir bénéficié de l'aide de ses enseignants dans la rédaction de ce projet. Enfin, le registre de langue qu'elle utilise est relativement soutenu et montre une familiarité avec le champ lexical de la sociologie. Lisa cite aussi des références légitimes en expliquant s'être intéressée à la formation en particulier l'écoute « suite à l'écoute d'une émission sur France Culture ». Le vocabulaire mobilisé traduit la forte dotation en capital culturel de Lisa et de sa famille et l'accompagnement rapproché de ses parents dans la rédaction de la lettre.

# Conclusion

Dans l'ensemble, les analyses ont confirmé qu'il existait des différences d'aspirations scolaires entre les bacheliers selon leur origine sociale et leur filière de baccalauréat. L'article propose un nouvel éclairage sur les inégalités d'orientation vers le supérieur entre les filières générales et les filières technologiques et professionnelles en se fondant sur deux matériaux originaux.

La recherche a montré que face à l'injonction faite aux enseignants d'accompagner davantage les élèves dans leur orientation à moyens constants, ceux-ci disposent principalement de deux stratégies d'adaptation. Le ciblage des bons élèves est une stratégie qui se retrouve plus souvent en filières technologiques et professionnelles – où les enfants de classes populaires sont surreprésentés – tandis que les enseignants délèguent davantage aux familles dans les filières générales. Les données nationales de l'enquête TSS confortent ainsi les études localisées décrivant des pratiques d'encadrement liées à la filière du baccalauréat et au profil social des élèves. Nous enrichissons ici ces études en soulignant le poids différencié du niveau scolaire des lycéens dans les pratiques d'encadrement selon la série du baccalauréat et proposons des pistes d'interprétation de ce phénomène.

Le deuxième constat majeur est que ces différentes stratégies enseignantes ont des effets sur l'intériorisation des prescriptions scolaires par les élèves et leur restitution dans les projets motivés sur Parcoursup. Dans les cas où l'accompagnement est plus ciblé, les bons élèves des filières technologiques et professionnelles, intériorisent fortement les consignes et leur place dans la hiérarchie des filières. Ils insistent sur la cohérence de leur choix avec leur cursus antérieur ou avec un projet professionnel précis, quand les bacheliers généraux évoquent un intérêt pour le contenu de la discipline. Dans les filières générales, où l'accompagnement est délégué aux proches, les lycéens de familles favorisées évoquent des éléments de leurs trajectoires personnelles tandis que les élèves moins dotés mobilisent essentiellement des dispositions scolairement valorisées. L'analyse montre ainsi comment s'articulent la série du baccalauréat (socialisation scolaire) et l'origine sociale (socialisation familiale) dans l'expression des aspirations. À terme, les projets motivés, bien que décrits comme une expression personnelle des candidats, restent fortement structurés par les modalités d'encadrement dont les élèves sont l'objet.

Lors de la mise en place de Parcoursup, les débats ont principalement porté sur l'élargissement de la sélection sur dossier scolaire aux licences et ses effets sur la ségrégation sociale et scolaire dans l'enseignement supérieur. Dans cet article, nous abordons le sujet sous un angle différent en rappelant ce que l'inégale répartition des élèves doit à la construction de leurs aspirations en amont du tri des candidatures. Nous nous plaçons ici d'un seul côté du « guichet » (celui des lycéens), d'autant que les lettres de motivation ne sont pas systématiquement examinées par les commissions de sélection. Par ailleurs, certaines formations – situées en bas de l'échelle d'un enseignement fortement hiérarchisé – sollicitent (par le jeu des désistements sur le portail) la quasi-totalité des candidats, c'est le cas de la licence de sociologie considérée. Après avoir été séparés à la sortie du collège unique dans des classes ou des établissements différents, des élèves aux dispositions et projections scolaires contrastées (visibles sur le plan factoriel) se retrouvent donc en coprésence dans certaines filières. Les effets de cette déségrégation sur leur socialisation universitaire, sur la perception de leurs valeurs relatives dans les hiérarchies scolaires et sur leurs parcours dans un contexte de massification de l'enseignement supérieur constituent un autre chantier.

# Bibliographie